\documentclass[10pt]{iopart}
\usepackage{graphicx,rotating,subfigure} 
\usepackage{iopams}  
\usepackage{dsfont}

\newcommand{\be}{\begin{eqnarray}}
\newcommand{\ee}{\end{eqnarray}}
\def\refeq#1{(\ref{#1})}
\def\d{\mbox d}
\def\wt{\widetilde}
\def\nn{\nonumber}
\def\i{\int_{-\infty}^{\infty}}
\def\ip{\int_{0}^{\infty}}

\def\G{\Gamma}

\def\la{\lambda}
\def\g{\gamma}
\def\al{\alpha}

\def\l{\left}
\def\r{\right}
\def\te{\mbox{e}}

\begin{document}
\title[Boundary susceptibility in the open $XXZ$-chain]{Boundary susceptibility in the open $XXZ$-chain}
\author{Michael Bortz\dag\ and Jesko Sirker\ddag}
\address{\dag\ Bergische Universit\"at Wuppertal, Theoretische Physik, 42097
  Wuppertal, Germany}
\address{\ddag\ School of Physics, The University of New South Wales, Sydney
  2052, Australia and\\
Department of Physics and Astronomy, University of British Columbia,
Vancouver, B.C., Canada V6T 1Z1}
\begin{abstract}
  In the first part we calculate the boundary susceptibility $\chi_B$ in the
  open $XXZ$-chain at zero temperature and arbitrary magnetic field $h$ by
  Bethe ansatz. We present analytical results for the leading terms when
  $|h|\ll \al$, where $\al$ is a known scale, and a numerical solution for the
  entire range of fields. In the second part we calculate susceptibility
  profiles near the boundary at finite temperature $T$ numerically by using
  the density-matrix renormalization group for transfer matrices and
  analytically for $T\ll 1$ by field theoretical methods. Finally we compare
  $\chi_B$ at finite temperature with a low-temperature asymptotics which we
  obtain by combining our Bethe ansatz result with recent predictions from
  bosonization.
\end{abstract}
\section{Introduction}
\label{intro}
Even a single impurity can have a drastic effect on the low-energy properties
of a one-dimensional interacting electron system. One of the simplest examples
is an antiferromagnetic spin-1/2 chain with a non-magnetic impurity which cuts
the chain and leads to a system with essentially free boundaries.  Because
translational invariance is broken the one-point correlation function $\langle
S^z(r)\rangle$ is no longer independent from the site index $r$ and the local
susceptibility $\chi(r)$ acquires a nonzero alternating part
\cite{EggertAffleck95}. Furthermore, the asymptotic behavior of correlation
functions near such a boundary is no longer governed by the bulk critical
exponents but instead by so called boundary or surface critical exponents
\cite{Cardy84NPB,WangVoit}. It is interesting to consider the case when the
spin chain is not cut but instead one of the links is only slightly weaker.
\begin{eqnarray}
\label{I1}
H &=& J\sum_{r=1}^{N-1}\left(S^x_rS^x_{r+1}+S^y_rS^y_{r+1}+\Delta
  S^z_rS^z_{r+1}\right) \nonumber \\
&+& J'_1 \left(S^x_1 S^x_N+S^y_1 S^y_N\right) +J'_2\Delta S^z_1 S^z_N\label{eq1}
\end{eqnarray}
Here $J>0,\,J'_{1,2}>0$ and we have allowed for an $XXZ$-type anisotropy which
is described by the parameter 
\be \Delta=:\cos\gamma \mbox{ with }
0\leq\Delta\leq1\; \; (0\leq\g\leq \pi/2)\,\label{gamdef}.  
\ee 
Using the Jordan-Wigner transformation
one can also think about this system as a lattice model of spinless Fermions
with a repulsive density-density interaction
\begin{eqnarray}
\label{I2}
H &=& \frac{J}{2}\sum_{r=1}^{N-1}\left( c_r^\dagger c_{r+1} +
  h.c. \right)+\Delta J \sum_{r=1}^{N-1} \left(n_r
  -\frac{1}{2}\right)\left(n_{r+1} -\frac{1}{2}\right) \nonumber \\
&+& \frac{J'_1}{2}\left( c_1^\dagger c_{N} +
  h.c. \right) + \Delta J'_2  \left(n_1
  -\frac{1}{2}\right)\left(n_{N} -\frac{1}{2}\right) \; .
\end{eqnarray}
Therefore $J'_1=J-\delta J$ with $0<\delta J\ll J$ corresponds to a weakening of
the hopping amplitude whereas $J'_2=J-\delta J$ gives a weakened
density-density interaction along one bond. For all $\Delta$ both
perturbations have the same scaling dimension $x=K/2$ where
$K=\pi/(\pi-\gamma)$ is the Luttinger parameter \cite{EggertAffleck92}.
Weakening the hopping or the interaction along one bond is therefore always
relevant\footnote{For the free Fermion case $K=2$ the perturbation
  becomes marginal.}. Assuming that the open chain presents the
only stable fixed point one therefore expects that the physics at energies
below $T_K/J \sim (\delta J/J)^{1/x}$ is governed by the open $XXZ$-chain
\cite{KaneFisherPRB}. This is the motivation to consider in the following only
the open boundary condition (OBC) $J'_1=J'_2=0$.

In an open $XXZ$-chain the free boundaries induce corrections of order $1/N$
to the bulk limit\footnote{Note that in the periodic case no $1/N$ corrections
  exists and the first correction to the bulk limit is of order $1/N^2$ and
  determines the central charge \cite{BloeteCardy}.}. In particular a $1/N$-term in the
susceptibility is expected which we denote hereafter as {\it boundary
  susceptibility} $\chi_B(h,T)$.  From the scaling arguments given before it
follows that a long chain with a finite concentration of impurities is
effectively cut into pieces of finite length. Measurements of the
susceptibility on such a system will therefore reveal large contributions from
the boundaries. This has inspired a lot of theoretical work to actually
calculate these boundary corrections
\cite{deSaTsvelik,AsakawaSuzuki96a,AsakawaSuzuki96b,Fujimoto,FujimotoEggert,FurusakiHikihara}.
Very recently the leading contributions to the boundary susceptibility for
$h\ll1$ and $T\ll 1$ have been calculated by field theoretical methods
\cite{FujimotoEggert,FurusakiHikihara}. On the other hand it is also known
that the $XXZ$-chain with OBC is exactly solvable by Bethe ansatz (BA)
\cite{AlcarazBarber,Sklyanin}. For zero temperature, however, only the leading
functional dependence on $h$ for the isotropic case $\chi_B(h,T=0)\sim 1/h(\ln
h)^2$ has been calculated so far \cite{AsakawaSuzuki96b,Fujimoto,frah97}. For
finite temperatures de Sa and Tsvelik \cite{deSaTsvelik} have applied the
thermodynamic Bethe Ansatz (TBA) in the anisotropic case. Evaluating their TBA
equations and comparing with a numerical solution (see section \ref{FT}) we
have found that their results are wrong for all anisotropies. Even the free
Fermion case (see section \ref{FF}) is not reproduced correctly and there is
also disagreement with the field theoretical results by Fujimoto and Eggert
\cite{FujimotoEggert} and Furusaki and Hikihara \cite{FurusakiHikihara} at low
temperatures. Frahm and Zvyagin \cite{frah97} have treated the isotropic case with the same
TBA technique. Although at least the functional form for low temperatures is
correct, their results are not reliable for high temperatures \cite{zvy04}.
This raises the question if the TBA is applicable at all for OBC or at least
which modifications have to be incorporated compared to the well known case of
periodic boundary conditions (PBC).

Our paper is organized as follows: We start with the simple but instructive
free Fermion case and establish results for the boundary susceptibility both
as a function of $T$ and $h$ in section \ref{FF}. In section \ref{BA} we report
the Bethe ansatz solution for $T=0$ and anisotropy $0\leq\Delta\leq1$. We
present analytical results for the boundary susceptibility at $|h|\ll 1$ and a
numerical solution of the BA formulas for arbitrary $h$. In section \ref{FT}
we calculate susceptibility profiles near the boundary numerically by the
density-matrix renormalization group for transfer matrices (TMRG) and
analytically by field theory methods. We also compare our numerical results for
$\chi_B(h=0,T)$ with an analytical formula for $T\ll 1$ which we obtain by
combining our BA results from section \ref{BA} with recent results from
bosonization \cite{FujimotoEggert,FurusakiHikihara}. The last section presents
a summary and conclusions.
\section{Free spinless Fermions}
\label{FF}
\begin{figure}[!htp]
\begin{center}
\includegraphics*[width=0.7\columnwidth]{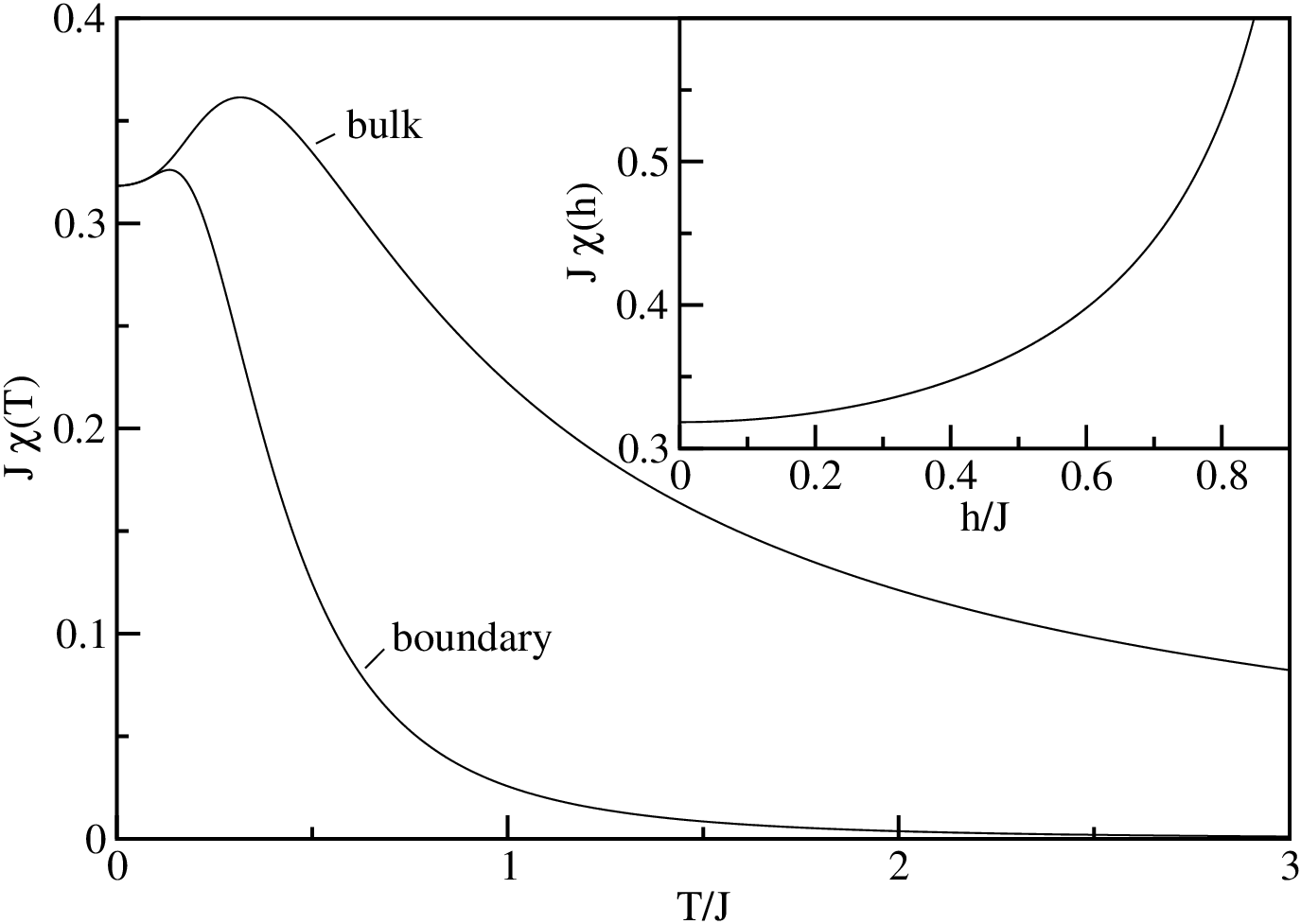}
\end{center}
\caption{Bulk and boundary susceptibilities for free Fermions. Note that
  $\chi(h,T=0)$ diverges for $h\rightarrow h_c = J$.}
\label{fig_FF1}
\end{figure}

Here we want to consider the special case $\Delta=0$ where eq.~(\ref{I2})
describes noninteracting spinless Fermions. After Fourier transform the
Hamiltonian takes the form
\begin{equation}
\label{FF1}
H = \sum_{n=1}^N \left(J\cos k_n +h\right)\left(c_{k_n}^\dagger c_{k_n}
  +h.c.\right)\quad\mbox{with}\quad k_n=\frac{\pi}{N+1}n
\end{equation}
where we have included a magnetic field $h$. Note that the only difference to
PBC are the momenta $k_n$ which in this case would be given by $k_n=2\pi/N$.
The susceptibility is easily obtained as
\begin{equation}
\label{FF2}
\chi(h,T) = \frac{1}{4T}\sum_n \cosh^{-2}\left[\frac{1}{2T}\left(J\cos k_n
    +h\right)\right]
\end{equation}
and using the Euler-MacLaurin formula then yields
\begin{eqnarray}
\label{FF3}
\chi(h,T) &=& \chi_{\mbox{\footnotesize bulk}}(h,T) +\frac{1}{N}\chi_B(h,T) +
\Or\left(\frac{1}{N^2}\right) \quad \mbox{with} \nonumber \\
\chi_{\mbox{\footnotesize bulk}}(h,T) &=& \frac{1}{4\pi T}\int_0^\pi \cosh^{-2}\left[\frac{1}{2T}\left(J\cos k
    +h\right)\right] dk \\
\chi_B(h,T) &=& \frac{1}{4\pi T}\int_0^\pi \cosh^{-2}\left[\frac{1}{2T}\left(J\cos k
    +h\right)\right] dk \nonumber \\
&-& \frac{1}{8T}\cosh^{-2}\left[\frac{1}{2T}\left(J+h\right)\right]
-\frac{1}{8T}\cosh^{-2}\left[\frac{1}{2T}\left(J-h\right)\right] \; . \nonumber
\end{eqnarray}
Therefore bulk and boundary susceptibility are identical at $T=0$ and given by
\begin{equation}
\label{FF4}
\chi_{\mbox{\footnotesize bulk}}(h,T=0) = \chi_B(h,T=0) =
\frac{1}{J\pi}\frac{1}{\sqrt{1-(h/J)^2}} \label{exff}\; .
\end{equation}
At finite temperatures $\chi_{\mbox{\footnotesize bulk}}$ and $\chi_B$ are
different, however, the additional factors in $\chi_B$ vanish exponentially
for $T\rightarrow 0$ so that they still share the same low-temperature
asymptotics
\begin{equation}
\label{FF5}
\chi_{\mbox{\footnotesize bulk}}(h=0,T\rightarrow 0) = \chi_B(h=0,T\rightarrow
0) = \frac{1}{J\pi}+\frac{\pi}{6J^3}T^2+\Or\l(T^4\r) .
\end{equation}
Fig.~\ref{fig_FF1} shows the boundary and bulk susceptibilities for finite
temperature at $h=0$ and as a function of $h$ at $T=0$ (inset).
In the next section we will discuss the BA solution for $T=0$. We will see
that a finite interaction between the Fermions, $\Delta\neq 0$, has dramatic
effects and $\chi_{\mbox{\footnotesize bulk}}(h,T=0)$ and $\chi_B(h,T=0)$ are no
longer identical. For $1/2\leq\Delta\leq 1$ we find that $\chi_B(h,T=0)$ even
diverges for $h\rightarrow 0$ whereas $\chi_{\mbox{\footnotesize bulk}}(h=0,T=0)$
remains always finite.
\section{The Bethe ansatz solution}
\label{BA}
In this section we calculate ground-state properties of the model \refeq{eq1}, i.e. $T=0$. The Hamiltonian \refeq{eq1} with $J_1'=J_2'=0$ has been diagonalized both by
coordinate and algebraic Bethe Ansatz \cite{AlcarazBarber,Sklyanin}. In the
following, we refer to the algebraic Bethe Ansatz \cite{Sklyanin}. The
eigenvalues $E$ are parameterized by a set of $M$-many quantum numbers
$\{\lambda_1,\ldots,\lambda_M\}$, \be E&=&J\l[-\sum_{j=1}^M\frac{\sin^2
  \gamma}{\cosh(2 \lambda_j)-\cos\gamma}
+\frac{N-1}{4}\,\cos\gamma\r]-h S^z\label{ej}\\
S^z&=&N/2-M\label{sz}.  \ee Here, $S^z$ is the total $z$-component of the
spin, $h$ a magnetic field along the $z$-direction and we always assume in the
following $h\geq 0$ without loss of generality. The anisotropic exchange
constant $\Delta$ is given by eq. \refeq{gamdef} and the $\lambda_k$ are
determined by the following set of coupled algebraic equations \be \fl
\frac{\phi(\la_k+\rmi \g/2)}{\phi(\la_k-\rmi\g/2)}\,\frac{a(2\la_k,\g)
  a(\la_k,\pi/2- \g/2)
  a(\la_k,\pi/2-\g/2)}{a(2\la_k,-\g) a(\la_k,-\pi/2+\g/2)a(\la_k,-\pi/2+\g/2)}\nn\\
\fl\qquad = -\frac{q_M(\la_k+\rmi\g) q_M(-\la_k-\rmi \g)}{q_M(\la_k-\rmi\g)
  q_M(-\la_k+\rmi \g)}\label{bae1}, \ee with the definitions \be \fl
\phi(\la):= \sinh^{2 N}(\la)\quad , \quad a(\la,\mu):=\sinh(\la+\rmi \mu)\quad
, \quad q_M(\la):= \prod_{j=1}^M \sinh(\la-\la_j)\nn\,.  \ee We first deal
with the anisotropic case $0<\g\leq \pi/2$ and obtain equations for the
susceptibility, which are solved analytically in the limit of small magnetic
field. The isotropic case is treated afterwards. At the end of this section,
we present numerical results for the susceptibility at arbitrary magnetic
field.

\subsection{Anisotropic case}
The solutions to \refeq{bae1} are periodic in the complex plane with period
$2\pi \rmi$, so that we can focus on a strip parallel to the real axis with
width $2\pi \rmi$. Using arguments of analyticity, one sees that there are
$2N+3+2M$ roots in such a strip. So besides the $M$-many $\la_k$ which yield
the energy eigenvalues \refeq{ej}, there are $2N+3+M$ additional roots. Consider
now the strip with Im $\la_k\in]-\pi,\pi] \, \forall k$. We denote the roots
in this set by
$\{\lambda_1,\ldots,\lambda_M,\lambda_1^{(h)},\ldots,\lambda^{(h)}_{2N+M},0,-\rmi\pi/2,\rmi\pi/2\}$.
It is straightforward to verify that $0,\pm \rmi\pi/2$ are roots. However, the
algebraic Bethe Ansatz fails for these roots so that these solutions must be
excluded. The roots are distributed symmetrically, both with respect to the
real and imaginary axis. In this work we focus on the calculation of the
ground state energy where $M=N/2$ and $\la_j\in \mathds{R}^{>0}\,\forall j$.
Then there are $N/2$ roots $\la^{(h)}_{j} =-\la_j$, $j=1,\ldots,N/2$. A
numerical evaluation of \refeq{bae1} shows that the remaining roots
$\lambda^{(h)}_{j=N/2+1,\ldots 3 N/2}$ ($\lambda^{(h)}_{j=3 N/2+1,\ldots
  5N/2}$) have imaginary part $-\rmi \g/2$ ($\rmi \g/2$). The eigenvalues $E$
are symmetrical in the $\la_j$. We thus want to deal with the set
$\{v_1,\ldots,v_N\}:=\{\la^{(h)}_{N/2},\ldots,\la^{(h)}_{1},\la_1,\ldots,\la_{N/2}\}$,
whose elements are distributed symmetrically on the real axis w.r.t. the
origin. From \refeq{bae1}, we find that the $v_j$ are the $N$ real solutions
to the equations \be \fl \frac{\phi(v_m+\rmi\g/2)}{\phi(v_m-\rmi
  \g/2)}\l[\frac{a(v_m,\pi/2-\g/2)\,a(v_m,\g/2)}{a(v_m,-\pi/2+\g/2)\,a(v_m,-\g/2)}\r]=\frac{q_N(v_m+\rmi\g)}{q_N(v_m-\rmi\g)}\label{bae2}.
\ee The remaining $2N$ solutions $v_j^{(h)}$ are identified as
$v_{j=1,\ldots,2N}^{(h)}\equiv \la^{(h)}_{j= N/2+1,\ldots,5 N/2}$. In
\refeq{bae2}, the terms in brackets $\l[\cdots\r]$ are due to the OBC. These
terms would be absent in the case of PBC.

Our aim is to calculate the $1/N$-contribution to the ground state energy per
lattice site in the thermodynamical limit (TL). Like in the PBC-case
\cite{kor93,tak99}, we introduce the density of roots on the real axis, 
\be
\Delta_{v_m}&:=&\frac{v_m+v_{m+1}}{2}-\frac{v_{m-1}+v_m}{2}\,,\qquad m=2,\ldots,N-1 \nn\\
\rho_+(v_m)&:=&\frac{1}{2N \,\Delta_{v_m}}\label{defrho}, 
\ee 
where $\Delta_{v_m}$ is the distance between two points on the left and on the right of the root $v_m$, such that the left (right) point is situated midway between $v_{m-1},v_m$ ($v_{m},v_{m+1}$). From numerical studies the boundary values of $\rho_+$ are inferred, $\rho_+(v_1)=\rho_+(v_N)=0$, with $\rho_+(v_{1,N}) \Delta_{v_{1,N}}=1$.  Together with \refeq{defrho} it follows that  
\be
\sum_{m=1}^N\rho_+(v_m)\Delta_{v_m}=\frac12\label{norm}\; .  
\ee 
In the TL
$\Delta_{v_m}\to0$ and $\rho_+(v_m)$ becomes a smooth function.  Let us define
the interval with non-vanishing density by $\l[-B,B\r]$, i.e.  $v_1\to-B$,
$v_N\to B$ in the TL. Then sums over functions $f(v_k)$ are transformed into
integrals by 
\be \sum_{k=1}^N\frac{1}{2N} f(v_k)=\int_{-B}^B f(x)\rho_+(x)\d
x-\frac{1}{2N}f(0)+\Or(1/N^2)\nn, 
\ee where the contribution $f(0)$ is
subtracted because the algebraic Bethe Ansatz fails at the origin.  There are
no further $\Or(1/N)$-terms because $\rho_+$ vanishes outside the integration
boundaries by definition,
\be
\rho_+(v)\equiv\rho_+(v)\theta(-v+B)\theta(B+v)\nn\; ,
\ee
where $\theta(v)$ is the Heaviside-function. In order to find the continuum version of \refeq{bae2}, it is convenient to define
\be
\rho(v):=\rho_+(v)+\rho_-(v)\nn\\
\rho_-(v):=\rho(v)(\theta(v-B)+\theta(-B-v))\nn\,. 
\ee
By taking the logarithmic derivative, the continuum version of
\refeq{bae2} is obtained \be 
\vartheta(x,\g)+\frac{1}{2N}\l[\vartheta(x,\g)+\vartheta(x,\pi-\g)+\vartheta(x,2\g)\r]\nn\\
\qquad=\rho(x)+\int_{-B}^B\vartheta(x-y,2\g)\rho_+(y)\,\d
y\label{int1}, \ee where \be 2\pi \rmi \,\vartheta(x,\g)&:=&\frac{2\rmi
  \sin\g}{\cosh 2x-\cos\g}=\frac{\d}{\d x}\ln \frac{\sinh(x+\rmi
  \g/2)}{\sinh(x-\rmi\g/2)}.  \ee Equation \refeq{int1} is a linear integral
equation with two unknowns, $B$ and $\rho$.  In a first step, \refeq{int1} is
solved for $B=\infty$; in a second step, $\rho_+(x)$ is obtained depending on
the parameter $B$ and the dependence of $B$ on the magnetic field $h$ is
calculated. We will see that $B=\infty$ corresponds to $h=0$, and a finite magnetic field $h>0$ induces a finite $B<\infty$. Finally, the susceptibility $\chi(h)$ is deduced. This procedure
was first used by Takahashi for PBC and is reviewed in \cite{tak99}.

Note that in deriving \refeq{int1} the range of definition of the involved
functions has been enlarged. All functions in \refeq{int1} are defined on
$[-\infty, \infty]$; to calculate physical quantities (like the ground-state
energy), however, we only need to know $\rho_+$, defined on $\l[-B,B\r]$.
Actually, it will be shown later that instead of dealing with $\rho_+$, all
quantities we are interested in can be expressed more conveniently by
$g_+(x):=\theta(x)\rho(v+B)$. The calculation of these functions is done by
Fourier transformation, \be \rho(x)&=&\frac{1}{2\pi} \i \widetilde
\rho(k)\te^{-\rmi kx}\,\d k\nn\; .  \ee Let us first consider the case
$B=\infty$, where $\rho\equiv \rho_+$. It is straightforward to solve
\refeq{int1} in Fourier space, where \be \widetilde
\vartheta(k,\gamma)&=&\frac{\sinh(\pi/2-\g/2)k}{\sinh\pi k/2} \nn.  \ee We
denote the solution of \refeq{int1} for $B=\infty$ by $\rho_0$ and find \be
\wt \rho_0(k)=\wt s(k)+\frac{1}{2N} \,\frac{\cosh\g k/4
  \,\cosh(\pi/4-\g/2)k}{\cosh\g k/2\,\cosh(\pi-\g)k/4}\label{rho0}, \ee with
\be \wt s(k):=\frac{1}{2\cosh\g k/2},\;s(x)=\frac{1}{2 \g \cosh\pi x/\g}\nn.
\ee Note that $\i\rho(x)\d x=1/2+1/(2N)$, in agreement with \refeq{norm}.

We now consider the case $B<\infty$, i.e., a finite magnetic field.  Let us derive the equation that determines $g_+$. Using
\refeq{rho0} we can rewrite \refeq{int1} as \be
\rho(x)&=&\rho_0(x)+\int_{|y|>B}\kappa(x-y)\rho(y)\,\d y\label{lie}\\
\kappa(x)&:=&\frac{1}{2\pi} \i \frac{\sinh(\pi/2-\g)k}{2\cosh \g
  k/2\,\sinh(\pi-\g)k/2}\,\te^{-\rmi kx}\,\d k\label{defkap}.  \ee We now
introduce the functions  
\be
\rho(x+B)=:g(x)\equiv g_+(x)+g_-(x)\nn\\
g_+(x)=\theta(x) g(x)\,,\qquad g_-(x)=\theta(-x) g(x)\label{defgpm}\,.
\ee
Then $g(x)$ satisfies the equation \be \fl
g(x)=\rho_0(x+B)+\ip \kappa(x-y)g(y)\,\d y+\ip \kappa(x+y+2B)g(y)\,\d
y\label{geq}.  \ee We seek a solution in the limit $B\gg 0$, which corresponds
to $|h|\ll \alpha$, where $\alpha$ is some finite scale which is calculated
later. The driving term $\rho_0(x+B)$ can be expanded in powers of
$\exp\l[B\r]$. Since \refeq{geq} is linear in $g$ and $\rho_0$, we make the
ansatz $g=g^{(1)}+g^{(2)}+\ldots$, where superscripts denote increasing powers
of $\exp\l[B\r]$. Then \be
\fl g^{(1)}(x)=\l[\rho_0(x+B)\r]^{(1)}+\ip\kappa(x-y)g^{(1)}(y)\,\d y\nn\\
\fl g^{(n)}(x)= \l[\rho_0(x+B)\r]^{(n)}+\ip \kappa(x+y+2B)g^{(n-1)}(y)\,\d
y+\ip \kappa(x-y)g^{(n)}(y)\,\d y\nn.  \ee Thus in each order, a linear
integral equation of Wiener-Hopf-type has to be solved. This technique is
explained for example in \cite{krei62,roo69}. It relies on the factorization
of the kernel $\wt \kappa$, \be 1-\tilde \kappa=1/(G_+ \,G_-)\label{gdef}, \ee
where $G_+$ ($G_-$) is analytical in the upper (lower) half plane and has
asymptotics $\lim_{|k|\to\infty}G_{\pm}(k)=1$. The functions $G_\pm$ are
calculated in \ref{appfac}. From \refeq{defgpm}, note that $\wt g_+$ ($\wt g_-$) is analytical in
the upper (lower) half of the complex plane. Then \be
\wt g_+^{(1)}(k)&=&G_+(k)\l[\wt \rho_0(k)G_-(k)\te^{-\rmi k B}\r]_+^{(1)}\label{g1}\\
\wt g_+^{(2)}(k)&=&G_+(k)\l\{\l[\wt \rho_0(k)\,G_-(k)\,\te^{-\rmi k
  B}\r]_+^{(2)}\r.\nn\\
& &\l.+\l[\wt \kappa(k)\,\wt g^{(1)}(-k)\,G_-(k)\,\te^{-2\rmi
  kB}\,G_-(k)\r]_+^{(2)}\r\}\,\label{g2}, \ee where \be
f_\pm(k):=\pm\frac{\rmi}{2\pi} \i \frac{f(q)}{k-q\pm\rmi \epsilon}\,\d
q\label{pint} \ee is analytical in the upper (subscript $+$) or lower
(subscript $-$) half of the complex plane such that $f=f_++f_-$. We will see later that it is sufficient to know $\wt g_+$. In this section we restrict
ourselves to the calculation of $\wt g_+^{(1)}$. The calculation of $\wt
g_+^{(2)}$ is sketched in \ref{ho}.

The bracket $\l[\ldots\r]_+^{(1)}$ in \refeq{g1} is evaluated using
\refeq{pint}, where only the pole nearest to the real axis is taken into
account. Then we find 
\numparts
\label{const1}
\be 
\fl \wt g_+^{(1)}(k)=\l\{\begin{array}{ll}
  G_+(k)\l\{\displaystyle\frac{a_0}{k+\rmi \pi/\g}\te^{-\pi/\g
    B}\r.& \\[0.4cm]
  \displaystyle\l.+\frac{1}{2N}\l[\frac{a_1}{k+\rmi\pi/\g}\te^{-\pi B/\g}+\frac{b_1}{k+\rmi
    2\pi/(\pi-\g)}\te^{-2\pi B/(\pi-\g)}\r]\r\}
\end{array}\r.\label{gp1}
\ee for $\g\neq\pi/3$ and \be \fl \wt g_+^{(1)}(k)=\displaystyle
G_+(k)\l[\frac{a_0}{k+\rmi 3 \pi}\te^{-3 B}+\frac{1}{2N}
\frac{c_1}{k+3\rmi}B\te^{-3B}\r] \label{gp11} \ee \endnumparts for $\g=\pi/3$ where the
constants are given by \numparts
\be
a_0&=&\frac{\rmi}{\g}G_-(-\rmi \pi/\g)\label{c1}\\
a_1&=& \frac{\sqrt{2} \rmi}{\g} G_-(-\rmi
\pi/\g)\frac{\sin\pi^2/(4\g)}{\cos(\pi^2/(4\g)-\pi/4)}\label{c2}\\
b_1&=& \frac{2\rmi}{\pi-\g}\tan\pi\g/(\pi-\g)\,G_-(-\rmi 2\pi/(\pi-\g))\label{c3}\\
c_1&=&\rmi \frac{18}{\pi^2} G_-(-3\rmi)\label{c4}.
\ee
\endnumparts
For $\g=\pi/3$, terms $\Or(B\exp\l[-B\r])$ occur in the boundary contribution which are absent for $\g=\pi/3$. These terms are leading compared to those $\Or\l(\exp\l[-B\r]\r)$ so that these latter have been neglected for the boundary contribution in \refeq{gp11}.

We are now ready to compute $s^z:=S^z/N$ and $e:=E/N$ from
\refeq{ej},\refeq{sz}: 
\be
\fl s^z=1/2-\int_{-B}^B\rho_+(x)\d x+1/(2N)\label{sz2}\\
\fl e=-h s^z-\frac{J\sin\g}{2}\int_{-B}^B \vartheta(x,\g)\rho_+(x)\d
x+\frac{J}{4}\l(\cos\g+\frac{2-\cos \g}{N}\r)\label{gs1}.  
\ee 
We insert
\refeq{lie} into \refeq{sz2} to obtain 
\be 
s^z=\frac{\pi}{\pi-\g}\, \wt
g_+(0)\label{sz3},  
\ee 
which is an exact statement, including all orders $\wt g^{(n)}$. 
It is convenient to calculate $e-e_0$, where
$e_0:=e(h=0)$. We use again \refeq{lie} which yields 
\be 
\fl e-e_0=-h
s^z+\frac{4J\pi\sin\g}{\g}\ip \frac{g_+(x)}{\cosh(x+B)\pi/\g}\,\d
x\nn\\
\fl \qquad =-\frac{h\pi}{\pi-\g} \l(\wt g^{(1)}_+(0)+\wt g^{(2)}_+(0)\r) + \frac{8\pi J \sin\g}{\g}\l[\l(
\wt
g^{(1)}_+(\rmi \pi/\g)+\wt g^{(2)}(\rmi \pi/\g) \r)\te^{-\pi B/\g}\r.\nn\\
\fl \qquad\qquad -\l.\wt g_+^{(1)}(3 \rmi \pi/\g) \te^{-3\pi B/\g}
+\Or\l(\te^{-3\pi B/\g}\wt g^{(2)}\r)\r]\label{gse}, 
\ee 
where in the last
equation we restrict ourselves to the given orders. Now $B$ is treated as a
variational parameter and is determined in such a way that \be
\frac{\partial}{\partial B}(e-e_0)=0\label{var}.  \ee In this section we
consider only the leading order in \refeq{gse}. Inserting \refeq{gp1},
\refeq{sz3}, \refeq{gse} in \refeq{var}, $B$ is obtained as a function of $h$,
\be
B&=&-\frac\g\pi \ln\frac h\al\label{bh1} \\
\al&:=& \frac{2\pi J \sin\g}{\g} (\pi-\g) \frac{G_+(\rmi
  \pi/\g)}{G_+(0)}\; . \label{alh} \ee Thus $\alpha$ sets the scale for $h$. The
restriction to the leading orders in $\exp[-B]$ is equivalent to the leading
orders in $h$ in the limit $|h|\ll \al$.

One now makes use of \refeq{alh} to determine $s^z(h)$ from \refeq{sz3}, and
therefrom $\chi(h)=\partial s^z/\partial h$. Inserting the explicit
expressions for $G_\pm$ from \ref{appfac} we find 
\numparts
\be
\fl\chi_{\mbox{\footnotesize bulk}}=\frac{\g}{(\pi-\g)\pi J
  \sin\g} \; .\label{chib}
\ee
The boundary contribution is given by
\be
\fl\chi_B(h)= \l\{\begin{array}{l} 
\displaystyle\frac{\g}{J(\pi-\g)\pi \sqrt{2}\,
  \sin\g}\,\frac{\sin\pi^2/(4\g)}{\cos(\pi^2/(4\g)-\pi/4)}\\[0.4cm]
\displaystyle+\frac{2\g\sqrt{\pi}}{(\pi-\g)^2}\tan\frac{\pi\g}{\pi-\g}\frac1\al\,\frac{\G\l(\pi/(\pi-\g)\r)}{\G\l(1/2+\g/(\pi-\g)\r)}\,(h/\al)^{-(\pi-3\g)/(\pi-\g)}\\
\end{array}
\r.\label{chi1} \ee for $\g\neq \pi/3$ with \be \fl \al= 2J (\pi-\g)\sqrt\pi\,
\frac{\sin \g}{\g}\, \frac{\G(1+\pi/(2\g))}{\G(1/2+\pi/(2\g))} \label{al} \ee
and by \be \fl\chi_B(h)= -\frac{1}{\sqrt{3}\pi^2}\l(\ln\frac{4\,h}{27\pi}+1\r)
\label{chi11} \ee \endnumparts for $\g=\pi/3$.  Note that the first term in
\refeq{chi1}, which is independent of magnetic field $h$, is the leading
contribution for $\g>\pi/3$ (pole closest to the real axis in \refeq{g1}). For
$\g<\pi/3$ the second term dominates and, in addition, this term is the
next-leading contribution for $\g>\pi/3$ (second pole in \refeq{g1}). For
$\pi/7<\g <\pi/3$ the constant term represents the next-leading contribution,
however, for $\g <\pi/7$ a pole at $6\pi\mbox{i} /(\pi-\g)$ in \refeq{g1} becomes
second nearest to the real axis and gives the next-leading contribution (see
\ref{ho}).

The second term in \refeq{chi1} is in perfect agreement with the result
obtained by conformal field theory and bosonization in
\cite{FujimotoEggert,FurusakiHikihara}. However, the first, field independent
term has not been obtained before. Also the result \refeq{chi11} for the
special case $\g = \pi/3$ is new. Our results are in qualitative agreement
with the TBA-work \cite{deSaTsvelik} at $T=0$, where also a finite value of
$\chi_B(T=0,h=0)$ for $\g>\pi/3$ (i.e. $\Delta<1/2$) and a divergent
contribution with the same exponent as in \refeq{chi1} for $\g<\pi/3$
($\Delta>1/2$) have been found. However, the coefficients in \refeq{chi1}
differ from those in \cite{deSaTsvelik}, due to an incorrect treatment of the
Bethe root at spectral parameter $x=0$ in \cite{deSaTsvelik} (c.f. the
discussion in section \ref{intro}). 
\subsection{Isotropic case}
\label{islim}
The isotropic case $\g=0$ (i.e. $\Delta=1$) is treated in the same manner as the
anisotropic case $\g\neq 0$. For the bulk susceptibility, \refeq{chib}, the
limit $\g\rightarrow 0$ can be performed directly, yielding
$\chi_{\mbox{\footnotesize bulk}}=1/J\pi^2$. For the boundary contribution
this limit is more complicated and we describe the procedure in the following. 

First, we rescale \refeq{ej} by $\la_j\to\g\la_j$. This is equivalent to
substituting $k\to k/\g$ in Fourier space. Then \be
\wt s(k)&=&\frac{1}{2\cosh k/2}\nn\\
\wt \rho_0(k)&=& \wt s(k)+\frac{1}{2N}\,\frac{1}{2\cosh k/2}
\l(1+\te^{-|k|/2}\r)\label{isrho0}\; .  \ee Whereas the analyticity properties
of the bulk contribution to \refeq{isrho0} are qualitatively the same as in
\refeq{rho0}, the boundary contribution shows, besides poles, a cut along the
imaginary axis. In \refeq{g1}, the $\l[\ldots\r]_+$-bracket thus yields
contributions $\Or(\exp\l[-const.\,B\r])$ from the poles, and algebraic
contributions due to the cut. The exponential contributions are clearly
sub-leading in comparison to the algebraic ones, so only the latter are
calculated in the following. Using eq. \refeq{facabs} from \ref{appfac} and
explicit expressions for $G_{\pm}(k)$, we find (omitting the bulk
contribution) \be \fl\wt g_+^{(1)}(k)=\l\{\begin{array}{ll}
  G_+(0)(\al_1/B+\al_2(\ln B)/B^2+\al_3/B^2)& \\[0.2cm]
  +\Or((\ln B)/B^3,1/B^3),& k=0\\[0.2cm]
  \rmi \al_1 G_+(k)/(k B^2),& k\neq 0
\end{array}
\r.\nn\\
\fl\al_1= \frac{1}{\sqrt 2 \pi}\,,\;\al_2= -\frac{\sqrt2}{4\pi^2}\,,\;\al_3= \frac{1}{\sqrt 2\pi^2}\l(\ln2-\frac{1}{2}\ln(2\pi)\r)\,.\nn
\ee
From \refeq{sz3}, \refeq{gse}, \refeq{var} we obtain
\be
B&=&-\frac1\pi \ln\frac h\al\label{bh2}\\
\al^{-1}&=& \frac{G_+(0)}{2\pi J G_+(\rmi \pi)}\; . \label{alh2}
\ee
These equations are obtained by those from the anisotropic case, \refeq{bh1},
\refeq{alh}, by scaling $B\to\g B$ and sending $\g\to 0$ afterwards. Carrying out the same steps which lead to
\refeq{chi1}, one finds the boundary contribution 
\be
\fl s^z_B(h)=-\frac14\l(\frac{1}{\ln h/h_0}+\frac{\ln|\ln h/h_0|}{2\ln
  ^2h/h_0}\r)+\mbox{o}(\ln^{-2}h) \label{szis}\\
\fl\chi_B(h)= \frac14\l(\frac{1}{h \ln ^2 h/h_0}+\frac{\ln|\ln
  h/h_0|}{h\ln^3h/h_0}-\frac{1}{2h\ln^3h/h_0}\r)+\mbox{o}\l(\frac{1}{h\ln^3h}\r)\label{chiis}\\
\fl h_0=\al/\sqrt{2}=J\pi \sqrt{\pi/\te}\; . \label{h0}
\ee
The scale $h_0$ has been chosen such that in \refeq{szis}, no terms
$\Or(\ln^{-2}h)$ occur. The results \refeq{chiis}, \refeq{h0} agree with the TBA-work by Frahm et
al.~\cite{frah97} for $T=0$. Furthermore, agreement is found with \cite{FujimotoEggert,FurusakiHikihara,AsakawaSuzuki96b}, where scales which differ from ours \refeq{h0} by a constant factor were used.  
\subsection{Numerical evaluation}
\begin{figure}[!htp]
\begin{center}
\includegraphics*[width=0.7\columnwidth]{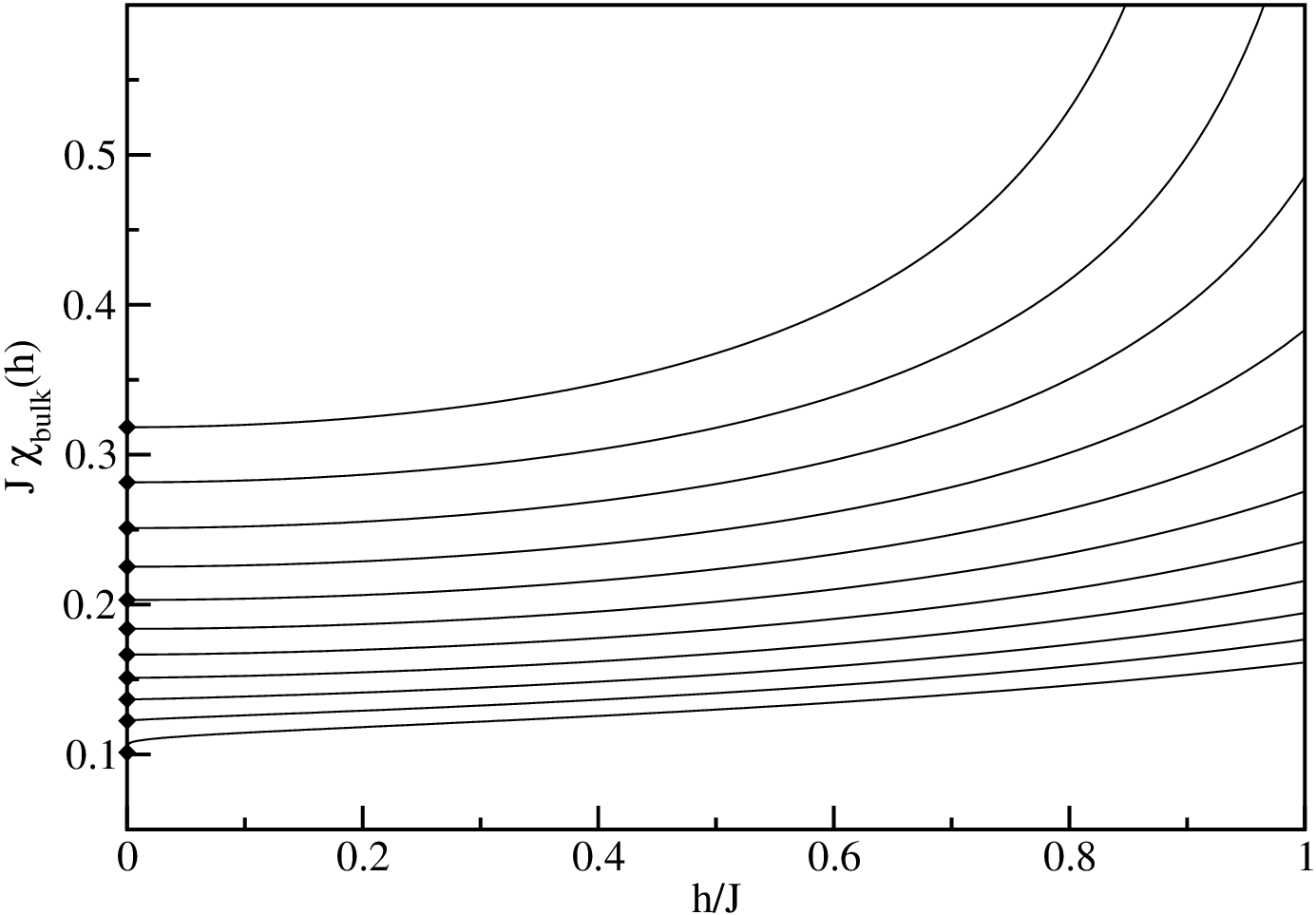}
\end{center}
\caption{Bulk susceptibility from a numerical solution of eq.~\refeq{int1}. The diamonds
  denote the $h=0$ values according to \refeq{chib}. Note that the $h=0$ value
  is approached with infinite slope in the isotropic case due to logarithmic
  terms, cf. eq. \refeq{chihis}.}
\label{fig_BA1}
\end{figure}

\begin{figure}[!htp]
\begin{center}
\includegraphics*[width=1.0\columnwidth]{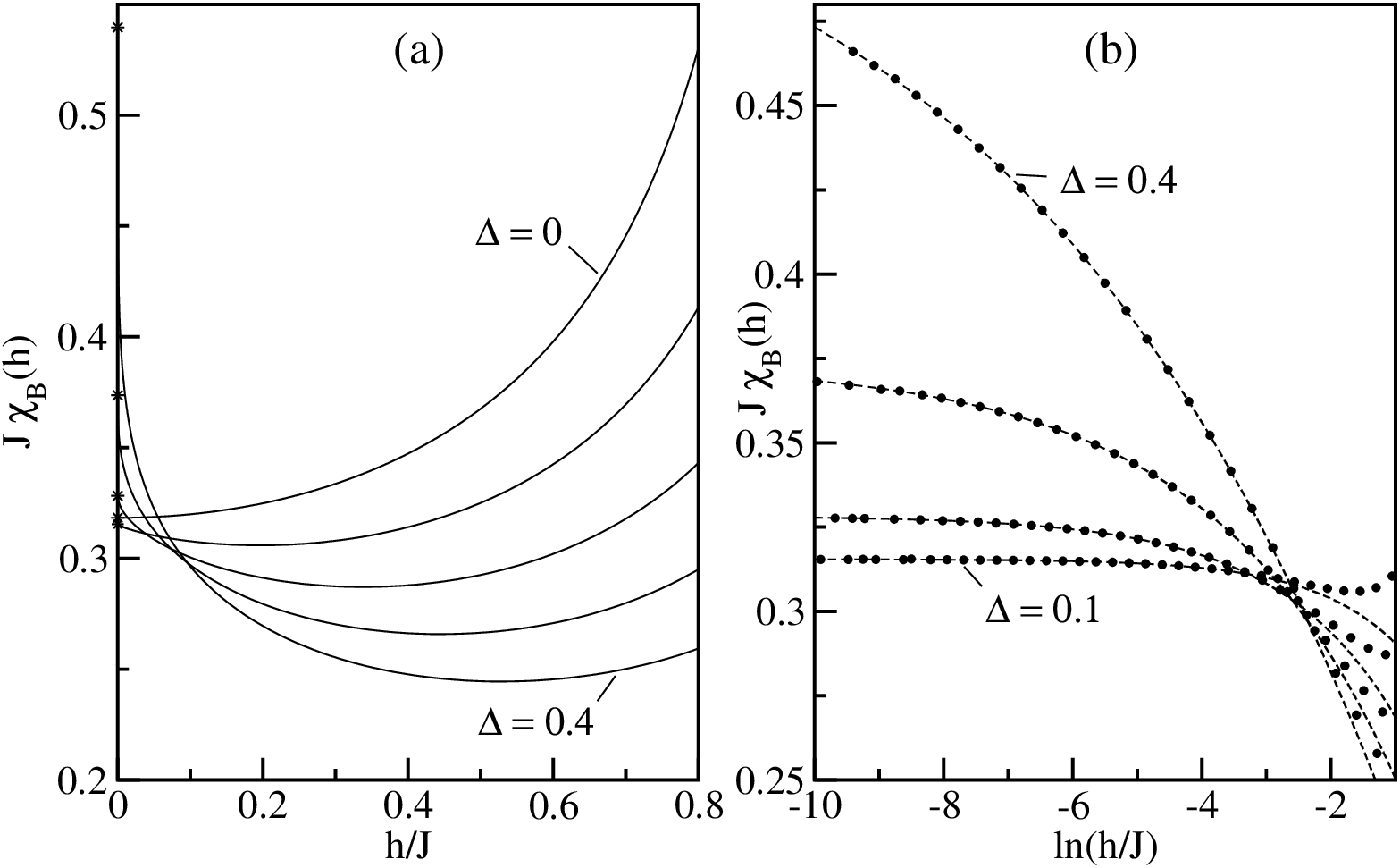}
\end{center}
\caption{(a) $\chi_B(h)$ for $\Delta=0.0,\cdots,0.4$. The stars denote the
  $h=0$ values according to eq.~\refeq{chi1}. (b) Comparison between the
  numerical solution (dots) of eq.~\refeq{int1} and the asymptotics (dashed
  lines) for $|h|\ll\alpha$ (eq.~\refeq{chi1}).}
\label{fig_BA2}
\end{figure}

\begin{figure}[!htp]
\begin{center}
\includegraphics*[width=1.0\columnwidth]{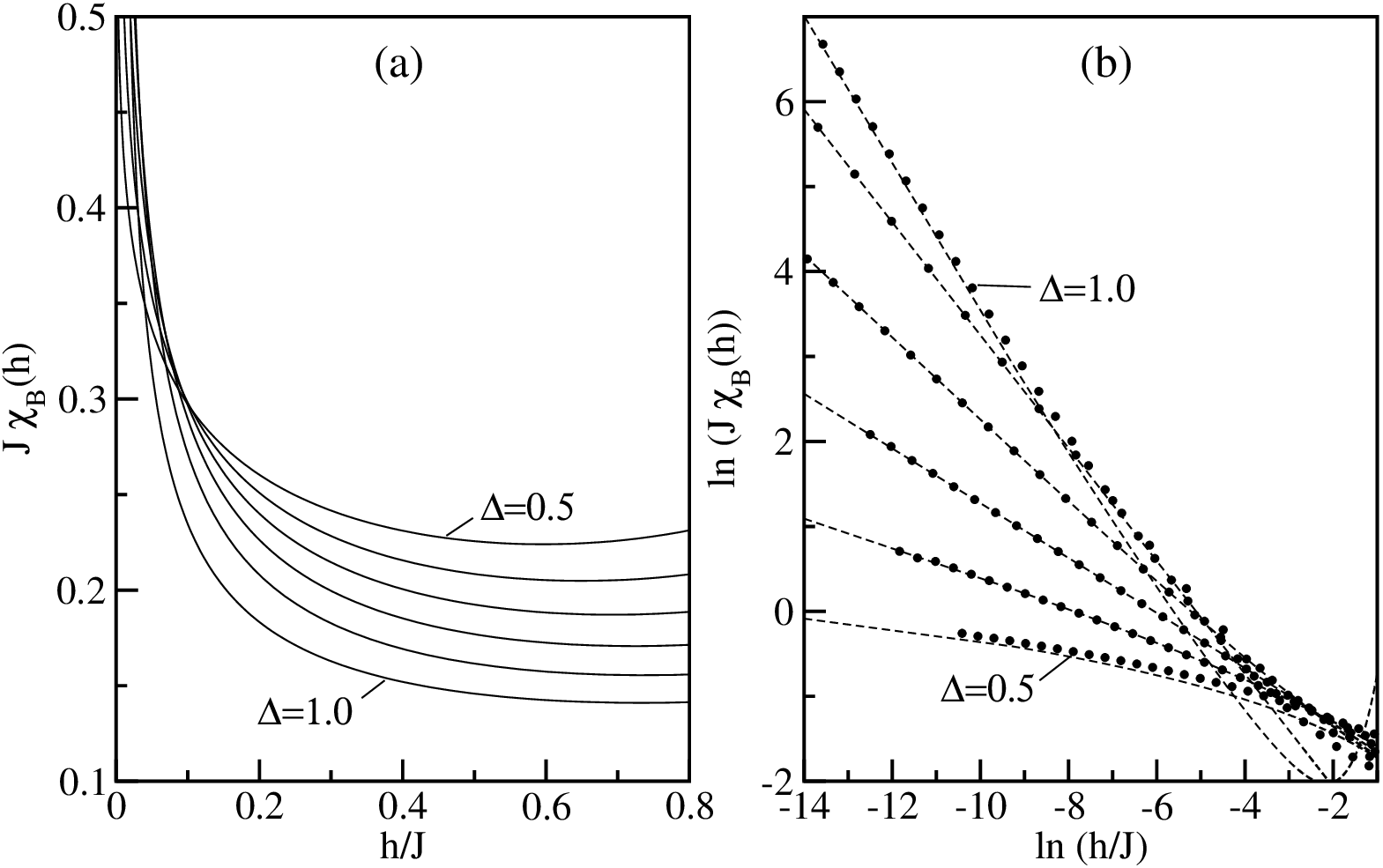}
\end{center}
\caption{(a) $\chi_B(h)$ for $\Delta=0.5,\cdots,1.0$. (b) Comparison between
  the numerical solution (dots) of eq.~\refeq{int1} and the asymptotics (dashed
  lines) for $|h|\ll\alpha$ (eq.~\refeq{chi1}).}
\label{fig_BA3}
\end{figure}

To obtain results for the case when $|h|\not\ll\alpha$, \refeq{int1} has to be
solved numerically. For this purpose the bulk and boundary contributions to
$\rho(x)$ in equation \refeq{int1} are treated separately. Both can be
evaluated numerically for arbitrary values of $B$. The corresponding value for
$h$ is then derived from the minimum condition \refeq{var}. This finally
yields $s^z(h)$ and therefrom $\chi(h)$, c.f. eq.~\refeq{sz2}. The result for the
bulk susceptibility is shown in fig. \ref{fig_BA1}, together with the $h=0$
values \refeq{chib}. The boundary susceptibility is shown in
fig.~\ref{fig_BA2} (\ref{fig_BA3}) for $\Delta<1/2$ ($\Delta>1/2$). In both
cases, the numerical solution confirms the analytical findings in the limit
$|h|\ll\alpha$.

\section{Finite temperatures}
\label{FT}
In the preceding section we calculated ground-state properties by making use
of the integrability of the Hamiltonian \refeq{eq1} with $J_1'=J_2'=0$. The
next step would be to exploit integrability in order to calculate
finite-temperature properties. However, the TBA seems to be problematic for
systems with OBC as discussed in the introduction. The other available
technique, namely the quantum-transfer-matrix-approach (QTM) \cite{kl93}, has
not been applied to open systems so far. The problem to modify the TBA
appropriately for OBC (if it is applicable at all) and the challenge to apply
the QTM-method for OBC remain open issues for future research. Here, we use
field theoretical arguments combined with a numerical study to discuss finite
temperatures.

First, we want to present a way different from section \ref{BA} to calculate the boundary
susceptibility. Because translational invariance is broken in a system with
OBC the one-point correlation function $\langle S^z(r)\rangle$ is no longer a
constant. The {\it excess magnetization} caused by the boundary can be defined
as
\begin{equation}
\label{eq_FT1}
M_{\mbox{exc}}(r) =  \langle S^z(r)\rangle_{\mbox{\footnotesize OBC}} -
M_{\mbox{\footnotesize PBC}}
\end{equation}
where $M_{\mbox{\footnotesize PBC}}$ is the magnetization per site in the
system with PBC and $r$ is the distance from the boundary. The local boundary
susceptibility is then given by
$\chi_B(r) = \partial M_{\mbox{exc}}(r)/\partial h$ and the total boundary
susceptibility $\chi_B$ can be obtained by
\begin{equation}
\label{eq_FT3}
\chi_B = \sum_{r=1}^{\infty} \chi_B(r) = \chi_{\mbox{\footnotesize OBC}} -
\chi_{\mbox{\footnotesize PBC}} \; .
\end{equation}
This means that we can calculate $\chi_B$ by considering only a local
quantity which is particularly useful in numerical calculations where it is
difficult to obtain the $1/N$ contribution directly. Particularly suited for
this purpose is the density-matrix renormalization group applied to transfer
matrices (TMRG) because the thermodynamic limit is performed exactly. The idea
of the TMRG is to express the partition function $Z$ of a one-dimensional
quantum model by that of an equivalent two-dimensional classical model which
can be derived by the Trotter-Suzuki formula \cite{Trotter,Suzuki2}. For the
classical model a suitable transfer matrix $T$ can be defined which allows for
the calculation of all thermodynamic quantities in the thermodynamic limit by
considering solely the largest eigenvalue of this transfer matrix. Details of
the algorithm can be found in
Refs.~\cite{BursillXiang,WangXiang,Shibata,SirkerKluemperEPL}. The method has
been extended to impurity problems in \cite{RommerEggert}. In particular, the local
magnetization at a distance $r$ from the boundary of a system with $N$ sites is given by
\begin{equation}
\label{eq_FT4}
\langle S^z(r)\rangle = \frac{\sum_n
  \langle\Psi_L^n|T(S^z)T^{r-1}\wt{T}T^{N-r-1}|\Psi_R^n\rangle}
{\sum_n \langle\Psi_L^n|T^{N-1}\wt{T}|\Psi_R^n\rangle}
\end{equation}
where $|\Psi_R^n\rangle$ ($\langle\Psi_L^n|$) are the right (left) eigenstates
of the transfer matrix $T$, $\wt{T}$ is a modified transfer matrix
containing the broken bond and $T(S^z)$ is the transfer matrix with the
operator $S^z$ included. Because the spectrum of $T$ has a gap between the
leading eigenvalue $\Lambda_0$ and the next-leading eigenvalues,
eq.~(\ref{eq_FT4}) reduces in the thermodynamic limit to
\begin{equation}
\label{eq_FT5}
\lim_{N\rightarrow\infty}\langle S^z(r)\rangle = \frac{\langle\Psi_L^0|T(S^z)T^{r-1}\wt{T}|\Psi_R^0\rangle}
{\Lambda_0^r \langle\Psi_L^0|\wt{T}|\Psi_R^0\rangle} \; .
\end{equation}
Therefore only the leading eigenvalue and the corresponding eigenvectors have
to be known to calculate the local magnetization in the thermodynamic limit.
Far away from the boundary $\langle S^z(r)\rangle$ becomes a constant, the
bulk magnetization
\begin{eqnarray}
\label{eq_FT6}
m= \lim_{r\rightarrow\infty}\lim_{N\rightarrow\infty}\langle S^z(r)\rangle 
&=& \lim_{r\rightarrow\infty}\frac{\sum_n\langle\Psi_L^0|T(S^z)T^{r-1}|\Psi_R^n\rangle\langle\Psi_L^n|\wt{T}|\Psi_R^0\rangle}
{\Lambda_0^r \langle\Psi_L^0|\wt{T}|\Psi_R^0\rangle} \nonumber \\
&=& \frac{\langle\Psi_L^0|T(S^z)|\Psi_R^0\rangle}{\Lambda_0}   \; .
\end{eqnarray}
To obtain numerically the susceptibility profile $\chi_B(r)$ at $h=0$ we
calculate $M_{\mbox{exc}}(r)$ for small fields $h/J\sim 10^{-2}$ to $10^{-3}$ by
using eqs.~(\ref{eq_FT5},\ref{eq_FT6}) and taking the numerical derivative. As
an example we show in fig.~5 
the susceptibility profile for $\Delta=0.6$ at various temperatures.


\begin{figure}[!htp]
\begin{center}
\includegraphics*[width=0.5\columnwidth]{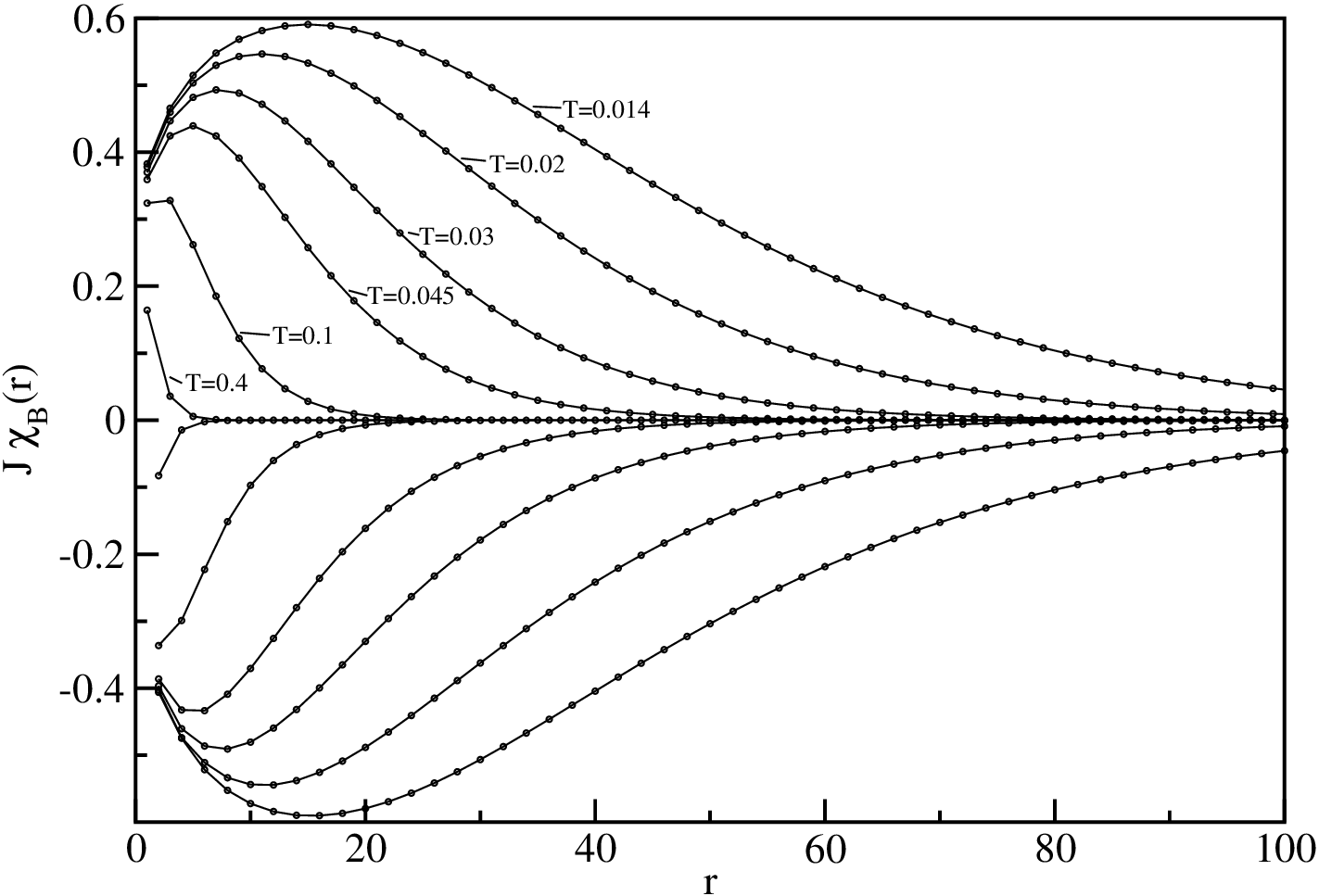}
\caption{Susceptibility profile for $\Delta=0.6$. The lines are a guide to the
  eye.}
\end{center}
\label{fig_FT0exp}
\end{figure}
\begin{figure}[!htp]
\includegraphics*[width=1.0\columnwidth]{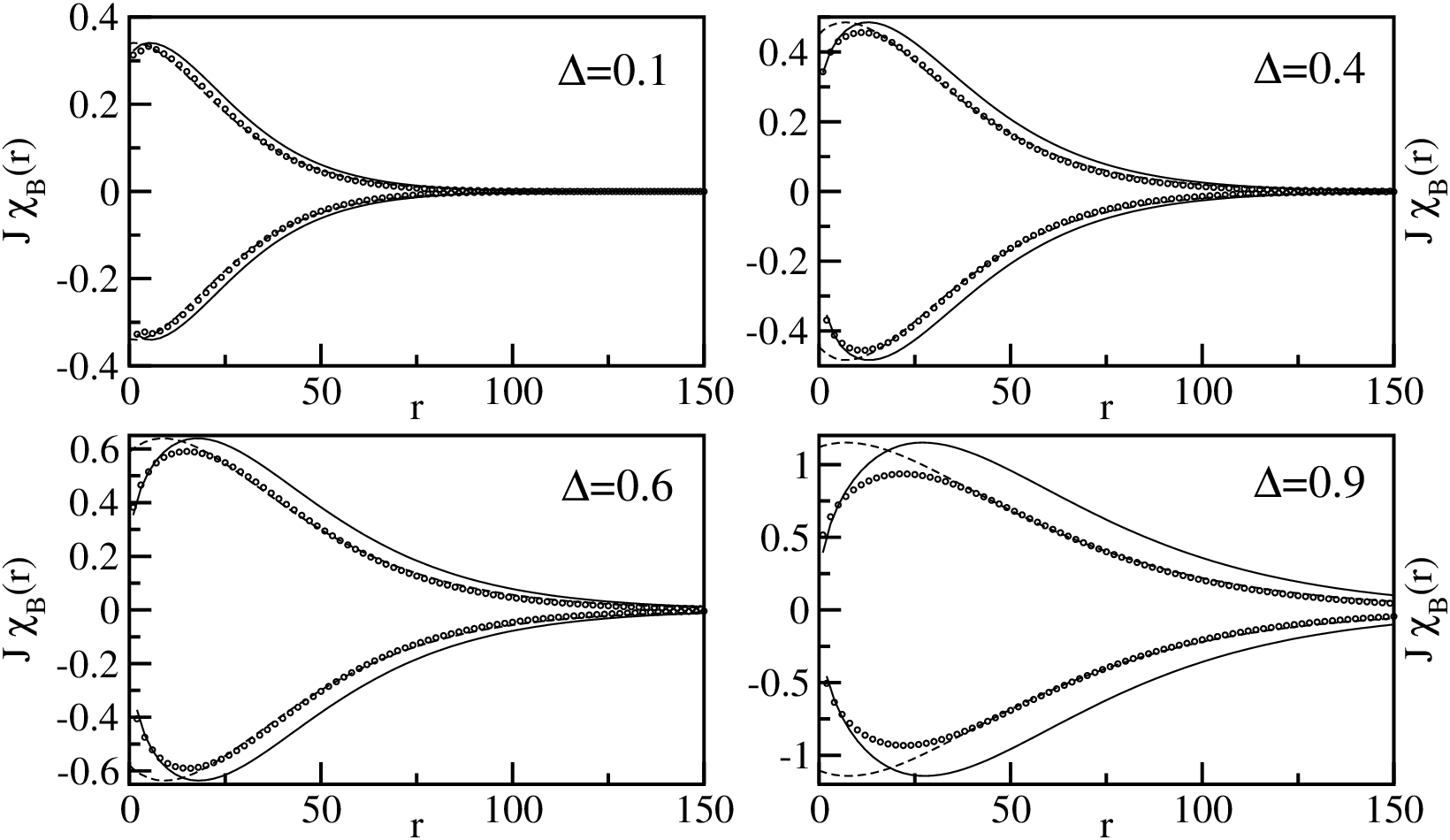}
\caption{Susceptibility profiles for different $\Delta$ at $T/J=0.01387$. The
  TMRG data are denoted by circles, the lines are the field theory results
  according to Eq.~(\ref{eq_FT11}) and the dashed lines the field theory
  results shifted by some lattice spacings (see text below).}
\label{fig_FT0}
\end{figure}

At sufficiently low temperatures the susceptibility profile exhibits a
maximum. This maximum gets shifted further away from the boundary as the
temperature is lowered. The dependence of $\chi_B(r)$ on $\Delta$ at fixed
temperature is shown in Fig.~\ref{fig_FT0}.
Here the maximum becomes more pronounced with increasing $\Delta$ and is at
the same time shifted further into the chain. 

Next we want to compare the numerics with field theory predictions. We start
with the bulk two-point correlation function $\langle S^z(r)S^z(0)\rangle$.
The leading term in the long distance asymptotics of this function at zero
temperature is known to be given by
\cite{LutherPeschel,WoynarovichEckle,BogoliubovIzergin,FrahmKorepin,LukyanovTerras}
\begin{equation}
\label{eq_FT7}
\langle S^z(r)S^z(0)\rangle \sim  A \frac{\cos(2k_F
  r+\phi)}{r^{2x}} 
\end{equation}
with an amplitude $A$ and phase $\phi$. The Fermi momentum is given by
$k_F=\pi(1\pm 2m)/2$ and the scaling dimension by $x=K/2$. The usual mapping
of the complex plane onto a semi-infinite cylinder then implies for small
temperatures
\begin{equation}
\label{eq_FT8}
\langle S^z(r)S^z(0)\rangle \sim  A \frac{\cos(2k_F
  r+\phi)}{\left(\frac{v_s}{\pi
      T}\sinh\frac{\pi T r}{v_s}\right)^K} \; . 
\end{equation}
Using Cardy's relation between $2n$-point functions in the bulk and $n$-point
functions near a surface \cite{Cardy84NPB} one can now directly obtain the
magnetization near the boundary 
\begin{equation}
\label{eq_FT9}
\langle S^z(r)\rangle \sim  \wt{A} \frac{\cos(2k_F
  r+\wt{\phi})}{\left(\frac{v_s}{\pi
      T}\sinh\frac{2\pi T r}{v_s}\right)^{K/2}} \; . 
\end{equation}
Note that although the critical exponent is only half the exponent appearing
in the two-point bulk correlation function both decay exponentially with
exactly the same correlation length $\xi = v_s/(2\pi x T)$. With the known
result for the bulk susceptibility $\chi_{\mbox{\footnotesize bulk}}(h=0) =
K/2\pi v_s$ we obtain
\begin{eqnarray}
\label{eq_FT10}
\langle S^z(r)\rangle &\sim&  \wt{A} \frac{(-1)^r\sin(Khr/v_s)}{\left(\frac{v_s}{\pi
      T}\sinh\frac{2\pi T r}{v_s}\right)^{K/2}} \quad \mbox{for} \; h\ll 1 \; \mbox{and} \nonumber \\
\chi_B(r)|_{h=0} &\sim&  \frac{\wt{A}K}{v_s} \frac{(-1)^r\, r}{\left(\frac{v_s}{\pi
      T}\sinh\frac{2 \pi T r}{v_s}\right)^{K/2}} \; .
\end{eqnarray}
This is the leading contribution to the boundary susceptibility. Note that
eq.~(\ref{eq_FT10}) agrees for the special case $\Delta=1$ with the result
given in \cite{EggertAffleck95}. The amplitude depends only on the operator
product expansion of $S^z$ and is given by $\wt{A}=\sqrt{A_z/2}$ where
$A_z$ has been derived by Lukyanov and Terras \cite{LukyanovTerras} (see
eq.~(4.3)). However, this alternating term does not contribute when
calculating $\chi_B$ by integrating over all lattice sites. The leading
non-oscillating contribution has already been obtained by Fujimoto and
Eggert \cite{FujimotoEggert} and Furusaki and Hikihara
\cite{FurusakiHikihara}. 
Including this term we find for the susceptibility profile
\begin{eqnarray}
\label{eq_FT11}
\chi_B(r)|_{h=0} &\sim&  \sqrt{\frac{A_z}{2}}\frac{K}{v_s} \frac{(-1)^r\, r}{\left(\frac{v_s}{\pi
      T}\sinh\frac{2 \pi T r}{v_s}\right)^{K/2}} \nonumber \\
&+& \frac{4K^2 \lambda}{v_s^2}\frac{r^2}{\left(\frac{v_s}{\pi
      T}\sinh\frac{2 \pi T r}{v_s}\right)^{2K}}
\end{eqnarray}
where the amplitude $\lambda$ is given in
\cite{LukyanovTerras,FurusakiHikihara}. This field theory result is shown as
straight lines in Fig.~\ref{fig_FT0} in comparison to the numerics. For all
$\Delta$ the shape of the curves agrees well with the numerical results.
However, especially at larger $\Delta$ the height of the maximum is
overestimated and there is also a shift by a few lattice sites. When we shift
the theoretical curves by an appropriate amount of lattice spacings (dashed
lines in Fig.~\ref{fig_FT0}) we see that the predicted exponential decay for
larger distances agrees perfectly with the numerical data. 

First, we should note that we cannot expect that the field theoretical
treatment yields reliable results for short distances. In addition, the
next-leading alternating terms in the asymptotic expansion of the bulk
two-point correlation function will become more and more important as
$\Delta\rightarrow 1$ \cite{LukyanovTerras}. This makes our approximation to
take only the leading term \refeq{eq_FT7} into account apparently worse for
larger $\Delta$. In fact shifting the field theoretical result is equivalent
to taking contributions with larger scaling dimensions into account. Our
observation that the shift increases with increasing $\Delta$ is therefore
consistent with the increasing importance of next-leading terms. We also want
to mention that a similar shift has been observed before for the isotropic
case \cite{EggertAffleck95}.

\begin{figure}[!htp]
\includegraphics*[width=1.0\columnwidth]{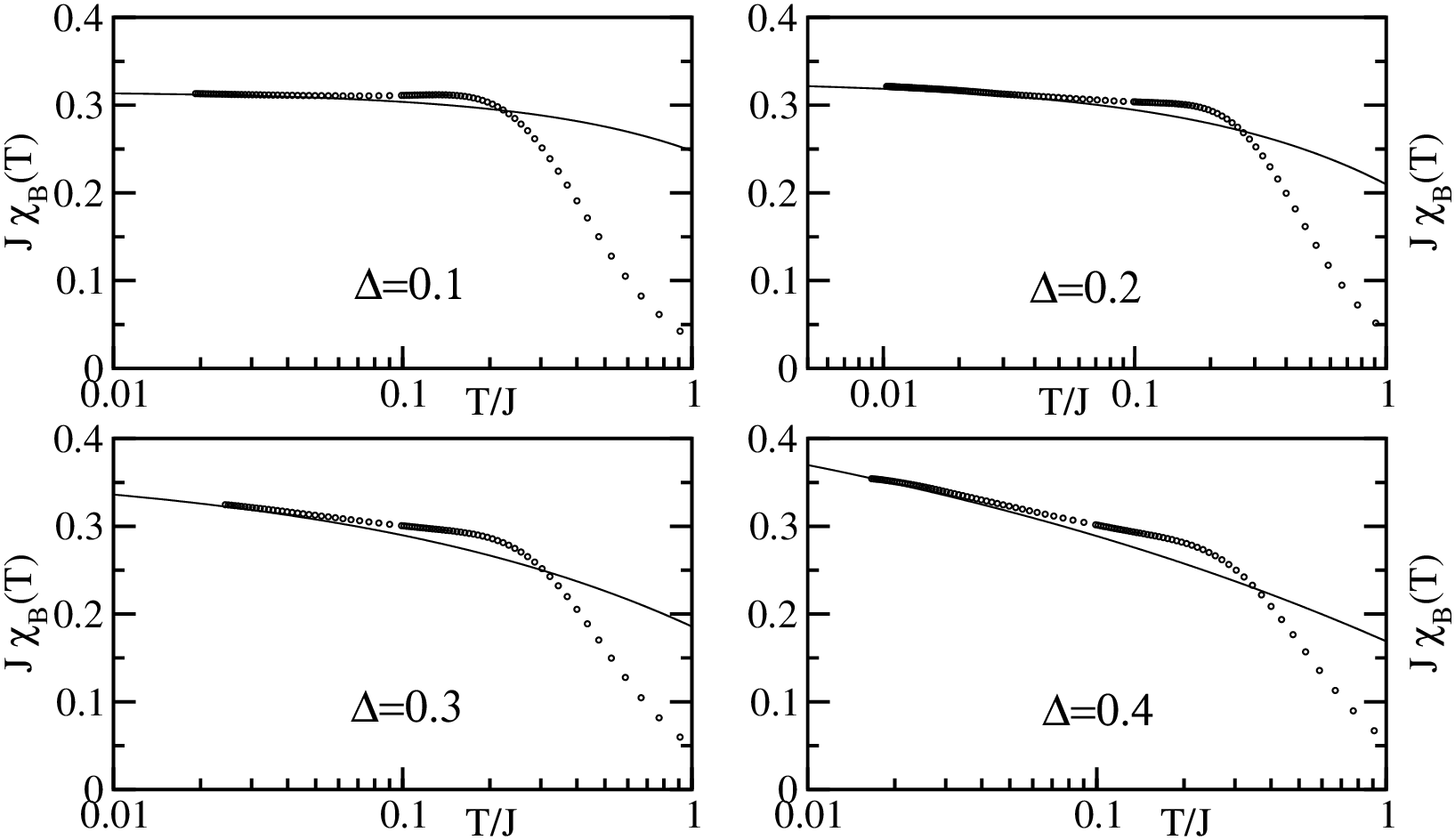}
\caption{Comparison between TMRG (circles) and the low-temperature asymptotics
  (lines) according to \refeq{eq_FT12} for $\Delta=0.1,\cdots,0.4$.}
\label{fig_FT1}
\end{figure}

\begin{figure}[!htp]
\includegraphics*[width=1.0\columnwidth]{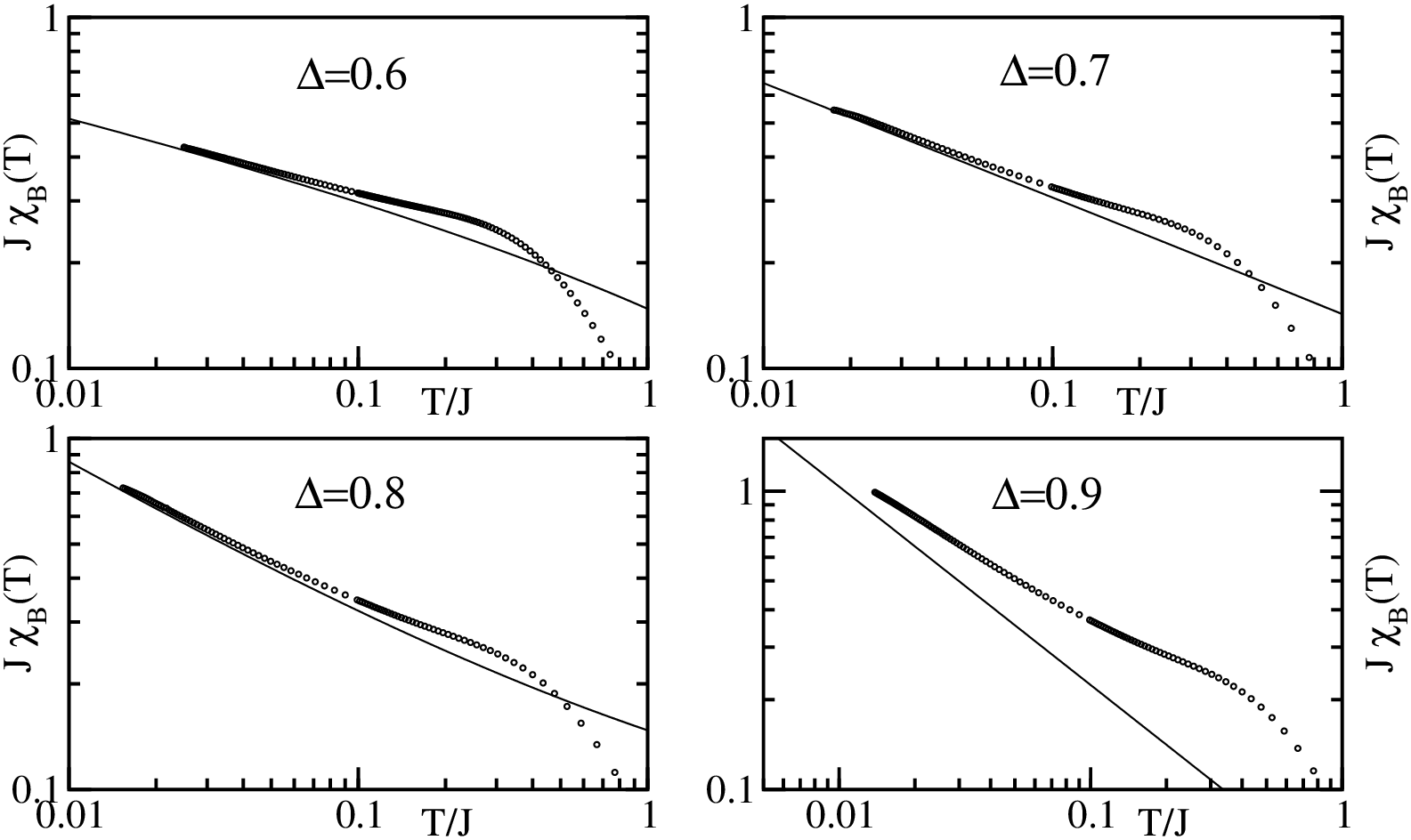}
\caption{Same as fig.~\ref{fig_FT1} with $\Delta=0.6,\cdots,0.9$.}
\label{fig_FT2}
\end{figure}

Finally we want to discuss the total boundary susceptibility $\chi_B$ at
finite temperature. To calculate it numerically we have in principle to sum
$\chi_B(r)$ over all lattice sites. However, at the lowest considered
temperature the correlation length $\xi<50$ and it is sufficient to take the
sum over the first 200 sites around the boundary. The results for
$\Delta=0.1-0.4$ are shown in Fig.~\ref{fig_FT1} and for $\Delta=0.6-0.9$ in
Fig.~\ref{fig_FT2}.
For the low-temperature asymptotics we already know from our Bethe ansatz
calculations that there is a temperature and magnetic field independent term
which dominates for $\Delta<0.5$.  In addition there is a temperature
dependent contribution which stems from the integral over all lattice sites of
the non-oscillating term in eq.~(\ref{eq_FT11}) and dominates for
$\Delta>0.5$. This integral has already been calculated in
\cite{FujimotoEggert,FurusakiHikihara}, yielding the leading
temperature-dependent contribution to $\chi_B(T)$. However, the constant
term cannot be calculated within the field-theoretical framework. Our knowledge of
this term from the BA calculations in the previous section allows us to obtain
a low-temperature asymptotics which is valid for all $0\leq\Delta < 1$. This
asymptotics is given by
\begin{eqnarray}
\label{eq_FT12}
\chi_B(T,h=0) &=&
\frac{K-1}{J \sqrt{2} \pi \sin(K-1)}\frac{\sin\frac{\pi}{4}\,\frac{K}{K-1}}{\cos \frac{\pi}{4}\,\frac{1}{K-1}}
\nonumber \\
&+& \lambda
\frac{K\Gamma(K)\Gamma(3-2K)(\pi^2-2\Psi'(K))}{4v^2(2-1/K)\Gamma(2-K)}\left(\frac{2\pi
    T}{v}\right)^{2K-3} \; ,
\end{eqnarray}
where $\Psi'(x)=\d^2\ln\G(x)/\d x^2$. The first term is our BA result,
eq.~\refeq{chi1}, where we have substituted $\g=\pi(K-1)/K$. The second term
is the leading temperature dependent contribution taken from
\cite{FujimotoEggert,FurusakiHikihara}. As in the case $h\neq 0$, $T=0$, we
expect that these are the leading and the sub-leading contributions for
$\Delta\lesssim 0.9$. For larger $\Delta$ we expect that other temperature
dependent terms will become more important than the constant term and we have
to omit it in this case to stay consistent. Note that the leading temperature
dependent term has the same exponent $2K-3$ as the leading magnetic field
dependent term in eq.~\refeq{chi1}. This is consistent with scaling arguments.
The lines in figs.~\ref{fig_FT1},\ref{fig_FT2} denote the low-$T$ asymptotics
according to this formula and are in excellent agreement with the numerics
except for $\Delta=0.9$ where the asymptotic limit at the lowest considered
$T$ is not yet reached. However, this is expected due to the afore mentioned
other temperature dependent terms which will become sub-leading for
$\Delta\gtrsim 0.9$ and even equally important as the term with exponent
$2K-3$ for $\Delta\rightarrow 1$, yielding finally a logarithmic dependence on
temperature \cite{FujimotoEggert,FurusakiHikihara}.
\section{Conclusions}
\label{end}
In the first part we have calculated the boundary contribution to the magnetic
susceptibility of the $XXZ$-chain with OBC at zero temperature and finite
magnetic field by BA. For small magnetic fields and $\g<\pi/3$ ($\Delta>1/2$) the BA result
for the leading divergent term agrees with the field theoretical analysis
\cite{FujimotoEggert,FurusakiHikihara}. For $\g>\pi/3$ ($\Delta<1/2$) a field independent
term is dominating. This term has not been obtained before. We also derived
for the first time the leading term for the special case $\g = \pi/3$
($\Delta=1/2$). In addition we have presented a numerical solution of the BA
equations for arbitrary field $h$. We used the numerics for a verification of
our analytical results for $|h|\ll\alpha$. 

In the second part we have calculated numerically susceptibility profiles near
the boundary by the TMRG method and compared these results with a
low-temperature asymptotics which we obtained by field theoretical methods.
Apart from a shift by a few lattice sites we have found good agreement. By
combining a temperature and magnetic field independent term, which we obtained
by BA and which is dominating for $\Delta<1/2$, with the leading temperature
dependent term, which has been calculated in
\cite{FujimotoEggert,FurusakiHikihara} and dominates for $\Delta>1/2$, we have
obtained a low-temperature asymptotics for $\chi_B(T)$ which is valid for all
$0\leq\Delta < 1$. Numerically, $\chi_B(T)$ has been obtained by a summation
of $\chi_B(r,T)$ over a sufficient number of sites around the boundary. At
low-temperatures excellent agreement with the analytical formula was found.
The remaining challenge is to calculate the finite-temperature properties
analytically by using the integrability of the model, either by TBA or by the
QTM-method.

We acknowledge very helpful discussions with Andreas Kl\"umper, Ian Affleck
and Frank G\"ohmann.

\appendix
\section{Factorization of the kernel}
\label{appfac}
Let us first carry out the factorization \refeq{gdef} for the anisotropic
case
\be
\frac{1}{G_+(k)\,G_-(k)}&=& \frac{\sinh\pi k/2}{2\cosh \g
  k/2\,\sinh(\pi-\g)k/2}\nn\\
G_-(k)&=&G_+(-k)\nn.
\ee
Using properties of the $\G$ function, we find
\be
G_+(k)&=&\sqrt{2(\pi-\g)}\frac{\G(1-\rmi k/2)}{\G(1/2-\rmi \g
  k/(2\pi))\,\G(1-\rmi k(\pi-\g)/(2\pi))}\,\te^{-\rmi a k}\nn\\
a&=& \frac12\l[\frac\g\pi\ln(\pi/\g-1)-\ln(1-\g/\pi)\r]\nn,
\ee
where $a$ is determined such that $\lim_{|k|\to\infty}G_{\pm}(k)=1$. 

As already mentioned in section \ref{islim}, the isotropic limit is realized
by scaling $k\to k/\g$, thus for $\g=0$, 
\be
\frac{1}{G_+(k)\,G_-(k)}&=&\frac{\te^{|k|/2}}{2\cosh k/2}\label{gis}.
\ee
By making use of 
\be
-\frac{|k|}{2}&=& \frac{\rmi k}{2\pi}\ln \rmi k-\frac{\rmi k}{2\pi} \ln(-\rmi
k),\label{facabs}
\ee
the exponential in \refeq{gis} is factorized in functions analytical in the
upper and lower half planes with
\be
G_+(k)&=& \sqrt{2\pi} \frac{(-\rmi k)^{-\rmi k/(2\pi)}}{\G(1/2+\rmi
  k/(2\pi))}\,\te^{-\rmi a k}\nn\\
a&=& -\frac{1}{2\pi}-\frac{\ln(2\pi)}{2\pi}\nn.
\ee
\section{Next-leading orders}
\label{ho}
Our focus here is on the anisotropic case; we comment on the
isotropic case at the end of this appendix. 

The calculation of the next-leading order, i.e. of $\tilde g^{(2)}$ in \refeq{g2}, is
technically more involved than the leading order $\tilde g^{(1)}$, because there are
{\em two} contributions in \refeq{g2}. The calculation is done following the
same steps as in section \ref{BA}, so that we merely give the results here. 

The $\l[\cdots\r]_+$-brackets in \refeq{g2} are evaluated using the integral
representation \refeq{pint}. Now, the pole next-nearest to the real axis is
taken into account in the first summand in \refeq{g2}. The second term
already contains a factor $\exp\l[-2\rmi k B\r]$, so that the pole next to the
real axis yields the leading contribution.  Thus we find for $\g\neq\pi/3$
\be
\fl \wt g_+^{(2)}(k)=G_+(k)\nn\\
\fl\qquad\times\l\{\l(\frac{a_{0,1}}{k+\rmi 3\pi/\g}+ \frac{a_{0,2}}{k+\rmi \pi/\g}\r)\te^{-3\pi
  B/\g}+\frac{a_{0,3}}{k+\rmi
  2\pi/(\pi-\g)}\,\te^{-(\pi/\g+4\pi/(\pi-\g))B}\r.\nn\\
\fl\qquad  +\frac{1}{2N}\l[\l(\frac{a_{1,1}}{k+\rmi
    3\pi/\g}+\frac{a_{1,2}}{k+\rmi\pi/\g}\r)\te^{-3\pi
    B/\g}+\frac{a_{1,3}}{k+\rmi
    2\pi/(\pi-\g)}\,\te^{-(\pi/\g+4\pi/(\pi-\g))B}\r.\nn\\
\fl\qquad+\l(\frac{b_{1,1}}{k+\rmi 6\pi/(\pi-\g)}\r.\nn\\
\fl\qquad\l.\l.\l.+\frac{b_{1,3}}{k+\rmi2\pi/(\pi-\g)}\r)\te^{-6\pi/(\pi-\g)B}
+\frac{b_{1,2}}{k+\rmi\pi/\g}\te^{-2(\pi/\g+\pi/(\pi-\g))B}\r]
\r\}\label{g2f}\\
\fl a_{0,1}=-\frac{\rmi}{\g} \,G_-\l(-\rmi \frac{3\pi}{\g}\r)\\
\fl a_{0,2}=
\frac{\rmi}{2\g\pi}\,\tan\frac{\pi^2}{2\g}\,G_-^3\l(-\rmi\frac\pi\g\r)\\
\fl a_{0,3}=
\frac{\rmi}{\pi(\pi+\g)}\,\tan\frac{\pi\g}{\pi-\g}\,G_-\l(-\rmi\frac\pi\g\r)\,G^2_-\l(-\rmi\frac{2\pi}{\pi-\g}\r)\\
\fl a_{1,1}= \rmi\frac{2\,\sin\pi/4\,\sin
  3\pi^2/(4\g)}{\g\,\cos(3\pi^2/(4\g)+\pi/4)}\,G_-\l(-\rmi
\frac{3\pi}{\g}\r)\\
\fl a_{1,2}= \frac{a_1}{2\pi} \tan\frac{\pi^2}{2\g}
G^2_-\l(-\rmi\frac\pi\g\r)\\
\fl a_{1,3}=
\frac{a_1\g}{\pi(\pi+\g)}\,\tan\frac{\pi\g}{\pi-\g}\,G_-^2\l(-\rmi\frac{2\pi}{\pi-\g}\r)\\
\fl b_{1,1}=\rmi\frac{2}{\pi-\g}\,\tan\frac{3\g\pi}{\pi-\g}\,G_-\l(-\rmi\frac{6\pi}{\pi-\g}\r)\\
\fl b_{1,2}=\frac{b_1(\pi-\g)}{\pi(\pi+\g)}\,\tan\frac{\pi^2}{2\g}\,G_-^2\l(-\rmi\frac\pi\g\r)\\
\fl b_{1,3}= \frac{b_1}{4\pi} \,\tan\frac{\pi\g}{\pi-\g}\,G_-^2\l(-\rmi\frac{2\pi}{\pi-\g}\r)\label{lasteq},
\ee
and for $\g=\pi/3$
\be
\fl\wt g_+^{(2)}(k)=G_+(k)\l\{\frac{a_{0,4}}{k+3 \rmi
  }\,B\,\te^{-9B}+\frac{1}{2N}\l[\frac{c_{1,1}}{k+9\rmi
      }B\te^{-9B}+\frac{c_{1,2}}{k+3\rmi }B^2\te^{-9 B}\r]\r\} \\
\fl a_{0,4}=\rmi\frac{9}{\pi^3}G_-^3(-3\rmi)\,,\;c_{1,1}= \rmi\frac{18}{\pi^2}\,G_-(-9\rmi)\,,\;c_{1,2}=\frac{3\,c_1}{\pi^2}\,G_-^2(-3\rmi)\nn.
\ee
This expression for $\wt g_2$ is inserted into \refeq{gse}, where we now have
to keep all the indicated terms.\footnote{It is indeed sufficient to restrict the
expansion of the $1/\cosh(x+B)$-factor in \refeq{gse} to the first two
orders. The next term would involve the exponent $5\pi/\g$. Comparing with the
largest exponent in \refeq{g2f}, $5\pi/\g>\pi/\g+4\pi/(\pi-\g)$ for
$\g<\pi/2$. However, $\g=\pi/2$ is allowed, since in this case, all
coefficients except $a_{0,1}$, $a_{1,1}$ vanish.} 
Then $B$ as a function of $h$ is derived. In section \ref{BA}, we found that
this relationship is the same both for the boundary and for the bulk in the
leading order. This is no
longer true when next-leading terms are considered. For the bulk we obtain
\be
\fl\al\te^{-\pi B/\g}=
h\l(1+A_1\l(\frac{h}{\al}\r)^2+A_2\l(\frac{h}{\al}\r)^{4\g/(\pi-\g)}+\l[\frac13\,\frac{a_{0,4}}{a_0}\l(\frac
h\al\r)^2\ln\frac h\al\r]_{\g=\pi/3}\r)\label{hb1}\\
\fl A_1= \frac{a_{0,2}}{a_0}+\frac{G_-(-\rmi 3\pi/\g)}{G_-(-\rmi\pi/\g)}\,,\;A_2= \frac{\pi-\g}{2\g}\,\frac{a_{0,3}}{a_0}\nn
\ee
and for the boundary
\be
\fl\al\te^{-\pi B/\g}=
h\l(1+A_1\l(\frac{h}{\al}\r)^2+A_2\l(\frac{h}{\al}\r)^{4\g/(\pi-\g)}+B_1\l(\frac{h}{\al}\r)^{-1+6\g/(\pi-\g)}\r.\nn\\
\fl\;\l.+B_2\l(\frac{h}{\al}\r)^{1+2\g/(\pi-\g)}+\l[\frac{c_{1,2}}{3 a_1} \l(\frac
h\al\r)^2\ln^2\frac h\al+\frac{c_1}{3}\,\frac{G_-(-9\rmi)}{G_-(-3\rmi)}\l(\frac
h\al\r)^2\ln\frac h\al\r]_{\g=\pi/3}\r)\label{hb2}\\
\fl A_1=\frac{a_{1,2}}{a_1}+\frac{G_-(-\rmi 3\pi/\g)}{G_-(-\rmi\pi/\g)}\,,\;A_2= \frac{\pi-\g}{2\g}\,\frac{a_{1,3}}{a_1}\nn\\
\fl B_1= \frac{2(\pi-\g)}{\pi+\g}\,\frac{b_{1,3}}{a_1}\,,\;B_2=\frac{\pi+\g}{2(\pi-\g)}\,\frac{b_{1,2}}{a_1}+\frac{G_-(-\rmi 3\pi/\g)}{G_-(-\rmi\pi/\g)}\,\frac{b_1}{a_1}\nn.
\ee
In \refeq{hb1}, \refeq{hb2} and in the following, for $\g=\pi/3$ the only nextleading terms are those labeled by $\l[\ldots\r]_{\g=\pi/3}$. Combining these equations with \refeq{sz3}, one finds
\be
\fl s_{\mbox{\footnotesize bulk}}^z(h)= \sqrt{\frac{2}{\pi(\pi-\g)}}\l\{G_-(-\rmi \pi/\g)\frac h\al \r.\nn\\
\fl\qquad+ \l(\frac{1}{\pi}\tan\frac{\pi^2}{2\g}\,G_-^3\l(-\rmi \frac\pi\g\r)-\frac13G_-\l(-\rmi \frac{3\pi}{\g}\r)+\frac{G_+(\rmi 3\pi/\g)G_+(\rmi \pi/\g)}{G_+(0)}\r)\l(\frac h\al\r)^3\nn\\
\fl\qquad +\frac{\pi-\g}{\pi(\pi+\g)}\,\tan\frac{\pi\g}{\pi-\g}\,G_-\l(-\rmi\frac\pi\g\r)\,G_-^2\l(-\rmi\frac{2\pi}{\pi-\g}\r)\,\l(\frac h\al\r)^{1+4\g/(\pi-\g)}\nn\\
\fl\qquad \l.-\l[\frac{2}{\pi^2}\,G^3_+(3\rmi)\l(\frac h\al\r)^3\ln\frac h\al\r]_{\g=\pi/3}\r\}\label{sbz}\\
\fl 2s^z_B(h)= -\rmi\sqrt{\frac{2}{\pi(\pi-\g)}}\l\{\g a_1\frac h\al+\frac{\pi-\g}{2}\,b_1\,\l(\frac h\al\r)^{2\g/(\pi-\g)}\r.\nn\\
\fl\qquad\l(2\g a_{1,2}+\frac{\g}{3}\,a_{1,1}+\g a_1\frac{G_+(\rmi 3\pi/\g)}{G_+(\rmi \pi/\g)}\r)\l(\frac h\al\r)^3\nn\\
\fl\qquad+\l((\pi-\g)a_{1,3}+\frac{\pi+\g}{2(\pi-\g)}\,\frac{b_1^2b_{1,2}}{a_1}+\g\frac{b_1^3}{a_1}\,\frac{G_+(\rmi 3\pi/g)}{G_+(\rmi \pi/g)}\r)\l(\frac h\al\r)^{1+4\g/(\pi-\g)}\nn\\
\fl\qquad+\l(\frac{2\g(\pi-\g)}{\pi+\g}b_{1,3}+\frac{\pi-\g}{6}b_{1,1}+\frac{\pi-\g}{2}b_{1,3}+\frac{\pi-\g}{2}\,\frac{b_1a_{1,3}}{a_1}\r)\l(\frac h\al\r)^{6\g/(\pi-\g)}\nn\\
\fl\qquad+\l(\frac{\g(3\pi-\g)}{2(\pi-\g)}b_{1,2}+\frac{2\g^2}{\pi-\g}\frac{b_1a_{1,2}}{a_1}+\frac{\g(\pi+\g)}{\pi-\g} \frac{G_+(\rmi 3\pi/\g)}{G_+(\rmi \pi/\g)}\r)\l(\frac h\al\r)^{2+2\g/(\pi-\g)}\nn\\
\fl\qquad+\frac{2\g(\pi-\g)}{\pi+\g}\frac{b_{1,3}}{a_1}\l(\frac h\al\r)^{8\g/(\pi-\g)-1}\nn\\
\fl\qquad+\pi\l[\frac{c_1}{9}\frac h\al \ln\frac h\al+\l(\frac{c_1c_{1,2}}{3^5 a_1}+\frac{c_{1,2}}{27}\r)\l(\frac h\al\r)^3\ln^2\frac h\al\r.\nn\\
\fl\qquad+\l.\l.\l(\frac{c_1^2}{9 a_1} \,\frac{G_+(9\rmi)}{G_+(3\rmi)}+\frac{c_{1,1}}{27\rmi}\r)\l(\frac h\al\r)^3\ln\frac h\al\r]_{\g=\pi/3}\r\}\label{sbbz}
\ee
From these expressions, $\chi$ can be obtained. Let us consider the special
case of free Fermions, i.e., $\g=\pi/2$. Eq.~\refeq{defkap} implies $\kappa\equiv 0$ so that $G_+\equiv G_-\equiv 1$. Furthermore, from \refeq{c1}-\refeq{c4}, \refeq{g2f}-\refeq{lasteq}, the only non-vanishing coefficients are
\be
\fl a_0= \rmi\frac{2}{\pi}\,,\; a_1=  \rmi\frac4\pi\,,\;a_{0,1}=-\rmi\frac2\pi\,,\;  a_{1,1}=-\rmi \frac4\pi\nn,
\ee
so that
\be
\fl \wt g_+^{(1)}(k)+\wt g_+^{(2)}(k)=\rmi\frac2\pi\l(\frac{1}{k+2\rmi}\te^{-2B}-\frac{1}{k+6\rmi}\te^{-6B}\r)\l(1+\frac1N\r)+\mbox{o}\l(\te^{-6 B}\r)\nn.
\ee
Note that this is the expansion of $\rho_0(x+B)$ in Fourier space. Equations \refeq{hb1}, \refeq{hb2} are equivalent in the free-Fermion case and yield
\be
\te^{-2B}&=&\frac h\al\l(1+\l(\frac h\al\r)\r)+\mbox{o}\l(h^2\r)\nn.
\ee
Finally, the sum of \refeq{sbz}, \refeq{sbbz} can be simplified to
\be
s^z(h)&=&\frac2\pi\l(\frac h\al+\frac23\l(\frac h\al\r)^3\r)\l(1+\frac1N\r)+\mbox{o}\l(h^3\r)\nn\\
\chi(h)&=&\frac1\pi\l(1+\frac12h^2\r)\l(1+\frac1N\r)+\mbox{o}\l(h^2\r)\label{appff},
\ee
where we have set $\al=2J$ for $\g=\pi/2$. Equation  \refeq{appff} is in agreement with the exact result \refeq{exff}
within the first two orders. 

As far as the isotropic case $\g=0$ is concerned, note that 
\refeq{szis}, \refeq{chiis} include already next-leading terms for the boundary
magnetization and susceptibility. Logarithmic corrections to the finite bulk
susceptibility, eq. \refeq{chib} with $\g=0$, have been calculated by
BA-techniques for PBC \cite{yang}:
\be
\chi_{\mbox{\footnotesize{bulk}}}(h)&=&\frac{1}{J\pi^2}
\l[1+\frac{1}{2\ln(\wt h_0/h)}-\frac{\ln\ln(\wt h_0/h)}{4(\ln(\wt h_0/h))^2}\r]+\mbox{o}\l(\ln^{-2}(h)\r)\label{chihis},
\ee 
with $\wt h_0=\al \te^{-1/8} \pi^{-1/4}$. The scale $\wt h_0$ has been determined such that no terms $\Or\l(\ln^{-2}h\r)$ appear in \refeq{chihis}. 
 

\end{document}